\documentclass[a4paper,11pt]{article}
\usepackage{pos}
\usepackage[normalem]{ulem}

\title{Lattice artifacts proportional to the quark mass in the QCD running coupling}

\author[a,b,c]{Marios Costa}
\author*[a,d]{Demetrianos Gavriel}
\author[a]{Haralambos Panagopoulos}
\author[a]{Gregoris Spanoudes}

\affiliation[a]{Department of Physics, University of Cyprus, Nicosia, CY-1678, Cyprus}
\affiliation[b]{Department of Mechanical Engineering and Material Science and Engineering, Cyprus University of Technology, Limassol, CY-3036, Cyprus}
\affiliation[c]{Rinnoco Ltd, Limassol, CY-3047, Cyprus}
\affiliation[d]{Dipartimento di Scienze Matematiche, Fisiche e Informatiche, Universita di Parma and INFN, Gruppo Collegato di Parma, I-43100 Parma, Italy}

\emailAdd{marios.costa@cut.ac.cy}
\emailAdd{demetrianos.gavriel@unipr.it}
\emailAdd{panagopoulos.haris@ucy.ac.cy}
\emailAdd{spanoudes.gregoris@ucy.ac.cy}

\abstract{Discretization artifacts proportional to the quark mass can limit the precision of strong-coupling determinations in lattice QCD, especially in the presence of heavy quarks. In this work, we perform a lattice perturbative analysis of such ${\cal{O}}(a m)$ effects in the running coupling by computing its two-loop renormalization factor $Z_g$. Using the background field method together with clover-improved Wilson fermions and Symanzik-improved gauge actions, we determine the mass-dependent components of the relevant two-point Green’s functions and obtain the improvement coefficients needed to remove ${\cal{O}}(a m)$ artifacts in mass-independent renormalization schemes. Our results are presented for general values of the number of colors $N_c$, the number of quark flavors $N_f$, and the clover coefficient $c_{\rm sw}$, and satisfy all symmetry and consistency constraints. Numerical values are provided for widely used instances of the above gauge actions, allowing improved control of mass-related cutoff effects in high-precision determinations of the strong coupling constant from lattice QCD. Full derivations and extended numerical results can be found in Ref.~\cite{Costa:2025xej}.}

\FullConference{The 42nd International Symposium on Lattice Field Theory (LATTICE2025)\\
2-8 November 2025\\
Tata Institute of Fundamental Research, Mumbai, India\\}

\begin{document}
\maketitle

\section{Introduction and motivation}
\label{sec:introduction}

High-precision determinations of the strong coupling $\alpha_s$ rely heavily on lattice QCD~\cite{DallaBrida2020, DelDebbio:2021ryq}. Lattice calculations allow one to compute short-distance observables nonperturbatively and match them to continuum schemes, providing a reliable way to extract $\alpha_s$ with controlled errors across a wide range of energy scales.

In practical applications, such as studies involving heavy quarks or decoupling strategies for scale evolution, the dimensionless quantity $a m_q$ can become large. In this regime, mass-dependent discretization effects of ${\cal O}(a m)$ are enhanced and can no longer be neglected, potentially introducing systematic errors in the running and matching of the coupling.

In mass-independent renormalization schemes, these effects are typically absorbed into a redefined modified bare coupling, $\tilde{g}_0^2$, which compensates for the linear mass dependence~\cite{Luscher1996}. Its definition depends on the improvement coefficient $b_g(g_0^2)$, which ensures that renormalized quantities remain free of linear cutoff effects even when quark masses are non-zero.

At present, $b_g(g_0^2)$ is known only to one loop in perturbation theory~\cite{DallaBrida:2022eua}, leaving a non-negligible uncertainty in precision determinations of $\alpha_s$. The primary goal of this work is a two-loop determination of $b_g(g_0^2)$, a step essential for improving the accuracy of strong coupling extractions on the lattice. We obtain this improvement coefficient from the mass-dependent part of the renormalization factor $Z_g$, computed here using the background field method.

These proceedings summarize the calculation and the main results of this effort; a comprehensive derivation and extended numerical data are available in Ref.~\cite{Costa:2025xej}. The remainder of this paper is structured as follows: Sec.~\ref{sec:setup} details the background field framework and the lattice actions, while Sec.~\ref{sec:methodology} describes the Feynman diagrams and loop integration methods. We present the results for $b_g$ up to two loops in Sec.~\ref{sec:results}, and offer concluding remarks in Sec.~\ref{sec:conclusions}.

%------------------------------------------------------------------

\section{Theoretical Setup}
\label{sec:setup}

\subsection{Background Field Framework}

The determination of the improvement coefficient $b_g(g_0^2)$ requires the computation of the renormalization factor $Z_g$, which relates the bare coupling $g_0$ to the renormalized coupling $g$:
\begin{equation}
g_0 = Z_g(g_0^2, a\mu) \, g.
\end{equation}
In this work, we compute $Z_g$ using the Background Field (BF) technique \cite{Ellis1984, Luscher1995background} within the framework of lattice perturbation theory. 

On the lattice, the gauge link $U_\mu(x)$ is decomposed into a quantum field $Q_\mu(x)$ and an external background field $B_\mu(x)$:
\begin{equation} 
\begin{split}
    U_\mu (x) &= U^Q_{\mu} (x) U^B_{\mu} (x),  \\
    U^Q_{\mu} (x) &\equiv e^{i g_0 Q_{\mu} (x)},  \\
    U^B_{\mu} (x) &\equiv e^{i a B_{\mu} (x)} .
\end{split}
\end{equation} 

This formulation allows gauge transformations to be interpreted in two distinct ways: either as quantum gauge transformations, where the background field remains invariant while the quantum field carries the full transformation, or as background gauge transformations, where the fields transform according to the following relations:
\begin{equation}
\begin{split}
    {[U^Q_\mu (x)]}^{\Lambda} &= \Lambda (x) U^Q_\mu (x) \Lambda^{-1} (x)  \\
    {[U^B_\mu (x)]}^{\Lambda} &= \Lambda (x) U^B_\mu (x) \Lambda^{-1} (x + a \hat{\mu}) .
\end{split}
\label{eq:bf_transformations}
\end{equation}
Here, $\Lambda(x)$ is the gauge transformation at site $x$. Under this second interpretation, the background field behaves as a standard gauge field, whereas the quantum field is restricted to local transformations. This separation enables gauge fixing while preserving background gauge invariance. Importantly, this invariance constrains how the background field can be renormalized, ensuring that physical observables constructed from it maintain the correct transformation properties under gauge transformations.

These constraints on renormalization have concrete implications. Specifically, background gauge invariance imposes a nontrivial constraint on the renormalization factors \cite{Abbott1980}, leading to
\begin{equation}
Z_g = Z_B^{-1/2},
\end{equation}
where $Z_B$ denotes the renormalization factor of the background field. As a consequence, the renormalization of the coupling constant can be determined entirely from the two-point function of the background field, thereby avoiding the more complex computation of three-point functions.

When nonzero quark masses are included, however, the fermion propagators and vertices are modified, affecting $Z_B$ and consequently the determination of $Z_g$. This introduces mass-dependent $\mathcal{O}(a m)$ contributions and additional technical complexity.

\subsection{Lattice actions}

The action employed in our calculation is:
\begin{equation}
S = S_{\rm F} + S_{\rm G} + S_{\rm gf} + S_{\rm FP} + S_{\rm meas.} ,
\end{equation}
where the Faddeev-Popov ($S_{\rm FP}$) and measure ($S_{\rm meas.}$) contributions follow the standard BF definitions \cite{Ellis1984}. We adopt the background gauge transformations defined in Eq.~(\ref{eq:bf_transformations}), under which the background field is treated as external and does not enter the path integral. To render the path integral finite while preserving background gauge invariance, the gauge-fixing term $S_{\rm gf}$ is chosen accordingly; its explicit form is detailed in Ref.~\cite{Costa:2025xej}.

For the fermions, we use the clover-improved Wilson action \cite{Sheikholeslami1985}:
\begin{eqnarray}
S_{\rm F} &=& \sum_{f}\sum_{x} (4r+m_0)\bar{\psi}_{f}(x)\psi_f(x)\nonumber \\
&-& \frac{1}{2}\sum_{f}\sum_{x,\,\mu}\bigg{[}\bar{\psi}_{f}(x) \left( r - \gamma_\mu\right)
U_{x,\, x+\mu}\psi_f(x+{\mu}) 
+\bar{\psi}_f(x+{\mu})\left( r + \gamma_\mu\right)U_{x+\mu,\,x}\psi_{f}(x)\bigg{]}\nonumber \\
&-& \frac{1}{4}\,c_{\rm sw}\,\sum_{f}\sum_{x,\,\mu,\,\nu} \bar{\psi}_{f}(x)
\sigma_{\mu\nu} {\hat F}_{\mu\nu}(x) \psi_f(x).
\label{eq:fermion_action}
\end{eqnarray}
Here, the index $f$ denotes the flavor, and $\sigma_{\mu \nu} = [\gamma_{\mu}, \gamma_{\nu}]/2$. The clover coefficient, $c_{\rm sw}$, is treated as a free parameter for the remainder of this work, while the Wilson parameter is fixed at $r = 1$. We assume all flavors have a degenerate Lagrangian mass $m_0$, although extending to flavor-dependent masses is straightforward. In addition, the tensor $\hat{F}_{\mu\nu}$ corresponds to the clover discretization of the gluon field-strength tensor.

For the gauge sector $S_{G}$, we adopt the Symanzik-improvement scheme \cite{Horsley:2004mx}, which includes $1\times1$ plaquettes and $1\times2$ rectangles. The lattice gauge action is
\begin{equation}
S_{\rm G}=\frac{2}{g_0^2} \left[ c_0 \sum_{\rm plaq.} {\rm Re\,Tr\,}\{1-U_{\rm plaq.}\} 
+ c_1 \sum_{\rm rect.} {\rm Re \, Tr\,}\{1- U_{\rm rect.}\} \right] ,
\end{equation}
with normalization condition
\begin{equation}
c_0 + 8 c_1 = 1,
\end{equation}
to ensure the correct classical continuum limit. For notational simplicity, explicit factors of the lattice spacing $a$ are suppressed; they can be restored by dimensional analysis whenever required. The plaquette ($U_{\rm plaq}$) and rectangle ($U_{\rm rect}$) loop variables are given by:
\begin{align}
U_{\rm plaq.}(x;\mu,\nu) &= U_\mu(x)\, U_\nu(x+\hat\mu)\, 
U_\mu^\dagger(x+\hat\nu)\, U_\nu^\dagger(x), \\
U_{\rm rect.}(x;\mu,\nu) &= U_\mu(x)\, U_\mu(x+\hat\mu)\, 
U_\nu(x+2\hat\mu)\, U_\mu^\dagger(x+\hat\nu+\hat\mu)\,
U_\mu^\dagger(x+\hat\nu)\, U_\nu^\dagger(x).
\end{align}

For numerical evaluations, we consider several commonly used gluonic actions with their corresponding Symanzik coefficients: the Wilson plaquette action with $c_0 = 1$ and $c_1 = 0$, the tree-level Symanzik (TLS) action with $c_0 = 5/3$ and $c_1 = -1/12$, and the Iwasaki action with $c_0 = 3.648$ and $c_1 = -0.331$. These three gauge actions are widely used in high-precision lattice simulations and span a representative range of discretization choices, from unimproved to improved formulations. Studying all three allows us to quantify the sensitivity of $b_g$ to the choice of gluonic discretization.

%------------------------------------------------------------------
\section{Methodology}
\label{sec:methodology}

\subsection{Feynman diagrams}

\begin{figure}[htbp]
  \centering
  \includegraphics[width=0.5\textwidth]{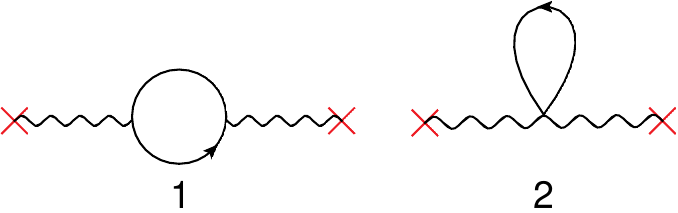}
  \caption{One-loop fermionic contributions to the background field two-point function. Solid lines denote fermions and wavy lines with a cross denote background gluons.}
  \label{fig:1loop}
\end{figure}

\begin{figure}[htbp]
  \centering
  \includegraphics[width=\textwidth]{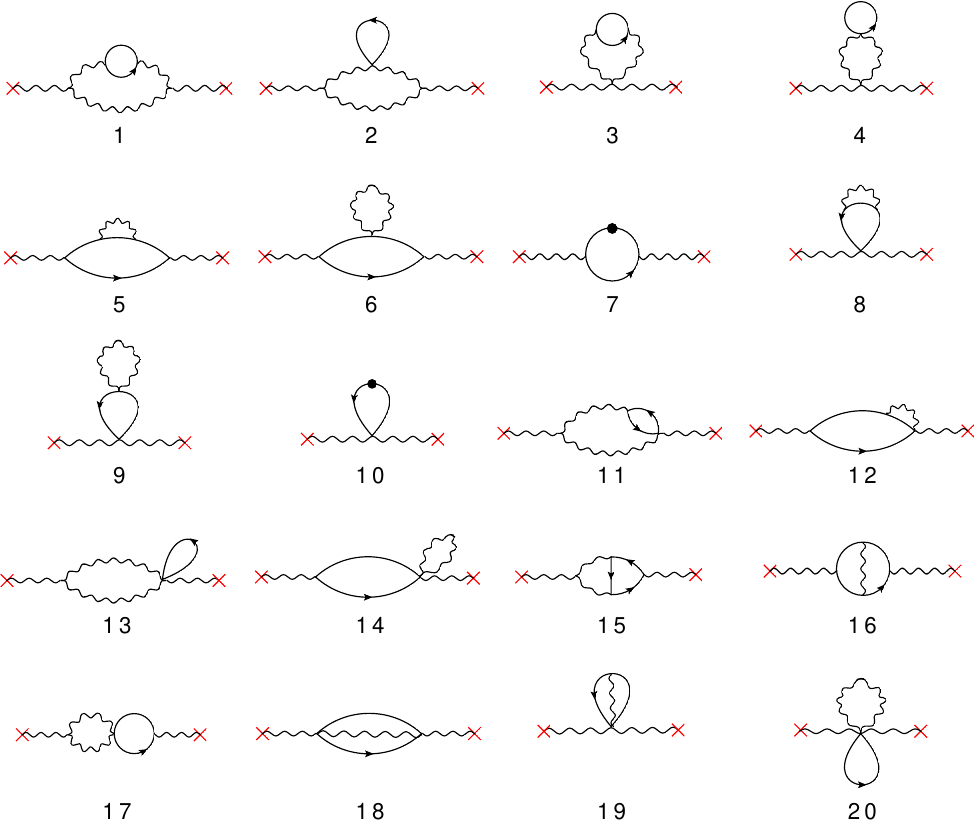}
  \caption{Two-loop fermionic contributions to the background field two-point function. Solid lines denote fermions, wavy lines represent gluons, wavy lines with a cross correspond to background gluons, and a solid line with a circle indicates the one-loop fermion mass counterterm.}
  \label{fig:2loop}
\end{figure}

The mass-dependent contributions to the renormalization of the coupling arise exclusively from Feynman diagrams containing at least one fermion line. In the background field formulation, they enter through the one-particle-irreducible two-point function of the background field evaluated at nonzero fermion mass. At one-loop order, the relevant diagrams consist of a single fermion loop coupled to two external background gluons, as shown in Fig.~\ref{fig:1loop}. At two-loop order, additional classes of diagrams appear, including those with internal gluon exchanges and insertions of the fermion mass counterterm. Diagrams are shown in Fig.~\ref{fig:2loop}. All diagrams are symmetrized with respect to Lorentz indices, color indices, and the external background field momenta. This reduces the number of independent tensor structures and provides a nontrivial check of the algebraic implementation.

\subsection{Loop momentum integration}

A central difficulty in the numerical evaluation of the loop integrals lies in the singular behavior of the integrands. Standard adaptive integration techniques are not directly applicable because the integrals contain pole structures. These originate from the massless gluon propagators and also from the fermion propagators once the latter are expanded in the fermion mass. The Taylor expansion introduces additional massless denominators, and in some cases enhances the degree of the pole.

If individual diagrams are examined in isolation, certain topologies exhibit severe infrared behavior. In particular, configurations involving two gluon propagators carrying identical momentum may generate poles of higher order, rendering those diagrams separately divergent. However, the underlying gauge symmetry enforces cancellations between specific diagrammatic contributions. When the appropriate combinations are formed, the singular structure softens to at most double poles. Since such singularities are integrable, the integrands are reorganized at the diagrammatic level prior to any numerical treatment.

After this, the momentum integrations are converted into finite sums by discretizing the Brillouin zone. The integration region for each loop momentum component is replaced by a uniform grid corresponding to a hypercubic lattice of finite extent. In the limit where the lattice size becomes arbitrarily large, the procedure approaches the exact infinite-lattice result. Points at which propagator denominators vanish are omitted from the summation; in fact, since they correspond to integrable singularities, their contribution will vanish in the infinite-lattice extrapolation. It is worth emphasizing that this construction does not coincide with finite-volume perturbation theory: the latter would require an explicit treatment of zero-momentum modes and their associated non-Gaussian functional contributions.

The discretization transforms the loop integrals into nested multi-dimensional summations containing a very large number of individual terms. Several measures are taken to reduce computational cost. Because the momentum variables assume only discrete values, trigonometric factors appearing in the integrands can be pre-evaluated in advance. Furthermore, symmetry properties, including reflections and reordering of summations, are exploited to restrict the evaluation to a reduced portion of the original summation domain. An additional optimization consists of reorganizing the algebraic expressions so that terms sharing common dependence on inner summation indices are grouped together. This hierarchical factorization avoids redundant evaluations in the most computationally demanding inner loops and significantly improves performance.

The entire workflow is automated: symbolic expressions are translated into efficient Fortran or C routines through a dedicated code-generation layer, enabling evaluations for a sequence of lattice sizes. Results obtained at finite lattice extent must subsequently be extrapolated to infinite volume. To control systematic uncertainties, several functional Ansätze describing the expected large-volume behavior are fitted to the numerical data. Their predictive quality is assessed by comparing each fit against lattice sizes not included in the determination of its parameters. Weighted averaging over the set of acceptable extrapolations yields the final infinite-volume estimate together with a systematic error. The reliability of this error assessment is supported both by agreement with analytically known cases and by the stability of results as additional, larger lattices are incorporated.

%------------------------------------------------------------------
\section{Two-loop results for the improvement coefficient}
\label{sec:results}

To eliminate $\mathcal{O}(am)$ lattice artifacts in mass-independent renormalization schemes, the bare coupling must be replaced by the improved coupling
\begin{equation}
\tilde{g}_0^2
= g_0^2 \left[
1 + a m_q
\left(
b_g^{(1)} g_0^2
+ b_g^{(2)} g_0^4
+ \mathcal{O}(g_0^6)
\right)
\right],
\label{eq:gtilde}
\end{equation}
where $m_q = m_0 - m_c(g_0^2)$ denotes the subtracted quark mass, with $m_c(g_0^2)$ defined as the critical mass at which the renormalized mass vanishes. This redefinition ensures that renormalized observables are free of linear mass-dependent cutoff effects. The coefficients $b_g^{(1)}$ and $b_g^{(2)}$ are determined from the mass-dependent part of the background field two-point function.

\subsection{One-loop coefficient}

At the one-loop level, the first-order coefficient $b_g^{(1)}$ does not depend on the number of colors $N_c$ or on the choice of gluon action; instead, it is determined solely by $N_f$ and the clover coefficient $c_{\rm sw}$. Specifically, this one-loop contribution is given by ($p$ is the external momentum):
\begin{equation}
b_g^{(1)} =
N_f \left[
0.0272837(1)
- 0.0223503(1)\, c_{\rm sw}
+ 0.0070667(1)\, c_{\rm sw}^2
- (1 - c_{\rm sw}) \frac{2}{16\pi^2}
\ln(a^2 p^2)
\right].
\label{eq:bg1}
\end{equation}
The logarithmic term in the above relation cancels once the perturbative expansion of $c_{\rm sw}$ is inserted, ensuring that the improved coupling is free of momentum dependence. In particular, setting the clover coefficient to its tree-level value, $c_{\rm sw}=1+\mathcal{O}(g_0^2)$, further simplifies the expression, yielding the well-known result originally derived in Ref.~\cite{Luscher1996}:
\begin{equation}
    b_g^{(1)} = 0.012000 N_f .
\end{equation}
The relatively small numerical value of this coefficient indicates that linear mass-dependent artifacts are suppressed at one loop, although higher-order terms could still provide non-negligible corrections.

\subsection{Two-loop coefficient}

We now turn to the computation of the two-loop coefficient $b_g^{(2)}$, which requires overcoming additional technical challenges compared to the one-loop case. The results for the Wilson, tree-level Symanzik (TLS), and Iwasaki gluon actions are summarized in Eqs.~(\ref{eq:bg2Wilson})–(\ref{eq:bg2Iwasaki}).
\begin{equation}
\begin{split}
    b&_g^{(2)}|_{\rm Wilson} = \frac{N_f}{N_c}  \Bigg\{  -0.0114014(29) - 0.015674(5) c_{\rm sw} + 0.0092076(20) c_{\rm sw}^2 \\
    & \hspace{2.65cm} + 0.00035230(5) c_{\rm sw}^3 - 0.000000475(4) c_{\rm sw}^4 + (1 - c_{\rm sw}) \left( \frac{\ln(a^2 p^2)}{16 \pi^2} \right)^2 \\
    & + \left( 0.0767570(8) + 0.1104702(14) c_{\rm sw} - 0.0408221(18) c_{\rm sw}^2 + 0.02123107(7) c_{\rm sw}^3 \right) \frac{\ln(a^2 p^2)}{16 \pi^2} \Bigg\} \\
    & \hspace{1.05cm} + N_f N_c \Bigg\{ -0.005463(5) + 0.003539(4) c_{\rm sw} - 0.0019859(13) c_{\rm sw}^2  \\
    & \hspace{2.65cm} - 0.00011429(12) c_{\rm sw}^3 + 0.000017174(4) c_{\rm sw}^4 + (1 - c_{\rm sw}) \left( \frac{\ln(a^2 p^2)}{16 \pi^2} \right)^2 \\
    & + \left( -0.156047(20) - 0.090502(20) c_{\rm sw} + 0.066449(5) c_{\rm sw}^2 - 0.0175896599(11) c_{\rm sw}^3 \right) \frac{\ln(a^2 p^2)}{16 \pi^2} \Bigg\}
\end{split}
\label{eq:bg2Wilson}
\end{equation}

\begin{equation}
\begin{split}
    b&_g^{(2)}|_{\rm TLS} = \frac{N_f}{N_c}  \Bigg\{  -0.0051236(24) - 0.013840(4) c_{\rm sw} + 0.0083484(19) c_{\rm sw}^2  \\
    & \hspace{2.65cm} + 0.00029588(7) c_{\rm sw}^3 - 0.000001975(4) c_{\rm sw}^4 + (1 - c_{\rm sw}) \left( \frac{\ln(a^2 p^2)}{16 \pi^2} \right)^2 \\
    & + \left( 0.0576153(9) + 0.0757940(14) c_{\rm sw} - 0.0367695(17) c_{\rm sw}^2 + 0.01954606(7) c_{\rm sw}^3 \right) \frac{\ln(a^2 p^2)}{16 \pi^2} \Bigg\}   \\
    & \hspace{1.05cm} + N_f N_c \Bigg\{ -0.007598(6) + 0.003048(6) c_{\rm sw} - 0.002202(5) c_{\rm sw}^2 \\
    & \hspace{2.65cm} - 0.00002829(13) c_{\rm sw}^3 + 0.0000156943(35) c_{\rm sw}^4 + (1 - c_{\rm sw}) \left( \frac{\ln(a^2 p^2)}{16 \pi^2} \right)^2 \\
    & + \left( -0.130706(33) - 0.056392(19) c_{\rm sw} + 0.059879(5) c_{\rm sw}^2 - 0.0165255904(23) c_{\rm sw}^3 \right) \frac{\ln(a^2 p^2)}{16 \pi^2} \Bigg\}  
\end{split}
\label{eq:bg2TLS}
\end{equation}

\begin{equation}
\begin{split}
    b&_g^{(2)}|_{\rm Iwasaki} = \frac{N_f}{N_c}  \Bigg\{  0.0031399(23) - 0.0112507(35) c_{\rm sw} + 0.0071443(16) c_{\rm sw}^2  \\
    & \hspace{2.65cm} + 0.00021121(5) c_{\rm sw}^3 - 0.0000035152(33) c_{\rm sw}^4 + (1 - c_{\rm sw}) \left( \frac{\ln(a^2 p^2)}{16 \pi^2} \right)^2 \\
    & + \left( 0.0370497(14) + 0.0368072(15) c_{\rm sw} - 0.0295570(18) c_{\rm sw}^2 + 0.01625262(8) c_{\rm sw}^3 \right) \frac{\ln(a^2 p^2)}{16 \pi^2} \Bigg\} \\
    & \hspace{1.05cm} + N_f N_c \Bigg\{ -0.012841(8) + 0.001980(9) c_{\rm sw} - 0.002891(17) c_{\rm sw}^2  \\
    & \hspace{2.65cm} + 0.00013071(11) c_{\rm sw}^3 + 0.0000129615(31) c_{\rm sw}^4 + (1 - c_{\rm sw}) \left( \frac{\ln(a^2 p^2)}{16 \pi^2} \right)^2 \\
    & + \left( -0.09865(8) - 0.017291(22) c_{\rm sw} + 0.048057(5) c_{\rm sw}^2 - 0.01418348(5) c_{\rm sw}^3 \right) \frac{\ln(a^2 p^2)}{16 \pi^2} \Bigg\} 
\end{split}
\label{eq:bg2Iwasaki}
\end{equation}

At two-loop order, the improvement coefficient exhibits a non-trivial dependence on the choice of gauge action.

\subsection{Numerical results for common gauge actions}

Initial inspection of the two-loop coefficients in Eqs.~(\ref{eq:bg2Wilson})--(\ref{eq:bg2Iwasaki}) reveals an explicit momentum dependence arising from various non-trivial logarithmic terms. However, this dependence is resolved once the perturbative expansion for the clover coefficient,
\begin{equation}
c_{\rm sw} = 1 + g_0^2 c_{\rm sw}^{(1)} + \mathcal{O}(g_0^4),
\end{equation}
is substituted into the general expression for the improvement coefficient, $b_g = b_g^{(1)} g_0^2 + b_g^{(2)} g_0^4 + \mathcal{O}(g_0^6)$. To ensure the consistent cancellation of these logarithms at the $g_0^4$ level, the first-order coefficient $b_g^{(1)}$ requires the $c_{\rm sw}$ expansion up to $\mathcal{O}(g_0^2)$, whereas for $b_g^{(2)}$, only the tree-level value is necessary. The resulting cancellation of all external momentum dependence serves as a critical consistency check; it ensures that the improvement coefficient is purely local, maintaining compatibility with the standard $\mathcal{O}(a)$ improvement framework.

For $N_c = 3$, and using the known values of $c_{\rm sw}^{(1)}$ corresponding to the
Symanzik-improved gauge actions considered in this work
\cite{Luscher:1996vw,Aoki:1998qd}, we obtain the following expressions:
\begin{equation}
b_g =
\begin{cases}
N_f \left(
0.012000(1)\, g_0^2
- 0.020067(20)\, g_0^4
+ \mathcal{O}(g_0^6)
\right), & \text{Wilson}, \\[6pt]
N_f \left(
0.012000(1)\, g_0^2
- 0.025347(30)\, g_0^4
+ \mathcal{O}(g_0^6)
\right), & \text{TLS}, \\[6pt]
N_f \left(
0.012000(1)\, g_0^2
- 0.04201(6)\, g_0^4
+ \mathcal{O}(g_0^6)
\right), & \text{Iwasaki}.
\end{cases}
\label{eq:bgFinal}
\end{equation}
These results reveal that the two-loop corrections provide a substantial contribution relative to the first-order terms. Such sizable higher-order effects are consistent with existing nonperturbative determinations of $b_g$ \cite{DallaBrida:2023fpl}, further highlighting that mass-dependent discretization effects remain a significant source of systematic uncertainty. Furthermore, the increasing magnitude of the $g_0^4$ coefficient as one moves from the Wilson action to improved gauge actions demonstrates that these cutoff effects are highly sensitive to the specific gauge action. This sensitivity should be taken into account in precision studies of the running coupling.

%------------------------------------------------------------------
\section{Discussion and outlook}
\label{sec:conclusions}

We have presented the first two-loop determination of the mass-dependent improvement coefficient $b_g$ which controls $\mathcal{O}(am)$ effects in the QCD running coupling. This result extends previous one-loop analyses and thereby completes the perturbative determination of this improvement term at next-to-leading order.

Although perturbative improvement alone may not fully capture mass-dependent cutoff effects at current lattice spacings, the two-loop determination of $b_g$ presented here provides an important benchmark. It can serve as input for refined perturbative analyses, a reference for future nonperturbative investigations, and a practical tool for lattice simulations with Wilson-type fermions, helping reduce systematic uncertainties associated with heavy-quark masses. Combining this perturbative input with future nonperturbative determinations is expected to further enhance control over mass-dependent effects in lattice QCD simulations.

%------------------------------------------------------------------
\section*{Acknowledgements}

It is a pleasure to thank Mattia Dalla Brida for fruitful discussions. This project is implemented under the programme of social cohesion ``THALIA 2021--2027'', co-funded by the European Union through the Research and Innovation Foundation (RIF). D.G. acknowledges support from the Istituto Nazionale di Fisica Nucleare (INFN) and the research project (iniziativa specifica) \textit{QCDLAT}.

\bibliographystyle{JHEP}
\bibliography{references}

\end{document}